\documentclass[letterpaper]{article} 
\usepackage{aaai2027}  
\nocopyright
\usepackage[hyphens]{url}  
\usepackage{graphicx} 
\usepackage{natbib}  
\usepackage{caption} 
\usepackage{amsfonts} 
\usepackage{mathrsfs} 
\usepackage{algorithm}

\usepackage{graphicx}
\newcommand{\tabincell}[2]{\begin{tabular}{@{}#1@{}}#2\end{tabular}} 
\usepackage{colortbl}
\definecolor{inteins}{RGB}{128,179,255}
\definecolor{intzwei}{RGB}{42,127,255}
\definecolor{intdrei}{RGB}{0,85,212}
\definecolor{intvier}{RGB}{255,143,143}
\usepackage{algorithm}
\usepackage{algpseudocode}
\usepackage{amsmath}
\usepackage{color}
\usepackage{multirow}
\usepackage{subcaption}
\usepackage{tabularx}
\usepackage{booktabs}
\newcolumntype{Y}{>{\centering\arraybackslash}X}
\newcolumntype{C}[1]{>{\centering\arraybackslash}p{#1}}

\usepackage{newfloat}
\usepackage{listings}
\DeclareCaptionStyle{ruled}{labelfont=normalfont,labelsep=colon,strut=off} 
\floatstyle{ruled}
\newfloat{listing}{tb}{lst}{}
\floatname{listing}{Listing}

\usepackage{booktabs}

\title{Beyond Static Forecasting: Unleashing the Power of World Models for Mobile Traffic Extrapolation}
\author{
    Xiaoqian Qi\textsuperscript{\rm 1},
    Haoye Chai\textsuperscript{\rm 2},
    Yue Wang\textsuperscript{\rm 1},
    Yong Li\textsuperscript{\rm 1}
}
\affiliations{
    \textsuperscript{\rm 1}Department of Electronic Engineering, BNRist, Tsinghua University, Beijing, China\\
    \textsuperscript{\rm 2}State Key Laboratory of Networking and Switching Technology, Beijing University of Posts and Telecommunications, Beijing, China\\
    qixiaoqian24@mails.tsinghua.edu.cn, haoyechai@bupt.edu.cn,\\
    wangyue@tsinghua.edu.cn, liyong07@tsinghua.edu.cn
}

\begin{document}

\maketitle

\begin{abstract}
Mobile traffic prediction is a fundamental yet challenging problem for wireless network planning and optimization. Conventional models mainly learn static long-term temporal patterns and cannot capture the dynamics under network-parameter adjustments. Leveraging the advantage of world models in learning underlying dynamics, we propose \textbf{MobiWM}, a mobile network world model that treats cell traffic as states and antenna parameters as actions. MobiWM combines factorized spatio-temporal modelling with multimodal environmental context aligned through shared spatial semantics. Its learned action-state transitions enable iterative rollout over specified adjustment trajectories for counterfactual planning. Extensive experiments on massive variable-parameter mobile traffic datasets demonstrate that MobiWM outperforms baselines by at least 16.40\% on average. A model-based Actor-critic case study further demonstrates its potential as a learned surrogate for network optimization.
\end{abstract}

\section{Introduction} 
\label{sec:intro}
Mobile traffic prediction is central to wireless network planning and optimization, enabling operators to allocate radio resources, balance load, and schedule energy-saving before congestion occurs~\cite{zhang2019deep_survey,wang2024survey_cellular}. However, reliable forecasting remains challenging because mobile traffic exhibits complex spatio-temporal dynamics driven by heterogeneous urban activities, changing engineering configurations, and topological interactions among densely deployed base stations (BSs). As networks evolve toward 5G-Advanced and 6G, the growing number of network nodes and increasingly flexible configurations further amplify traffic variability and inter-cell dependencies. Consequently, prediction models must move beyond static historical patterns and capture how network parameter adjustments reshape future traffic distributions.

Existing approaches have made substantial progress in modelling the spatio-temporal patterns of mobile traffic~\cite{bai2020agcrn, yang2024fedgtp, bettouche2025histm, wang2025mobimixer}. These methods attempt to capture diverse spatio-temporal dynamics through tailored architectural designs~\cite{li2018dcrnn, yu2018stgcn, wu2019graphwavenet, bai2020agcrn}, as well as to incorporate complex environmental correlations by fusing external context~\cite{xu2022cartagenie, chai2025stkdiff}. As network topology has a strong influence on the network operation, graph-based methods such as FedGTP~\cite{yang2024fedgtp} exploit spatial dependencies across distributed BSs under privacy constraints. State-space architectures like HiSTM~\cite{bettouche2025histm} leverage hierarchical Mamba modules for efficient long-horizon cellular traffic forecasting. Meanwhile, emerging foundation-model paradigms, exemplified by MobiFM~\cite{mobifm2025}, attempt to unify heterogeneous mobile data types within a single pre-trained backbone, advancing scalability and generalization. 

Despite these advances, existing methods mainly learn fixed-period and regular static traffic patterns. In real networks, traffic is also influenced by parameter tuning that operators perform to optimize coverage and capacity. Changing antenna power, azimuth, or tilt may alter coverage footprints, trigger user handover, and redistribute traffic across neighboring cells. Current models cannot reliably answer what would happen after a hypothetical configuration change.
\begin{figure*}[tb]
\centering
\includegraphics[width=0.95\linewidth]{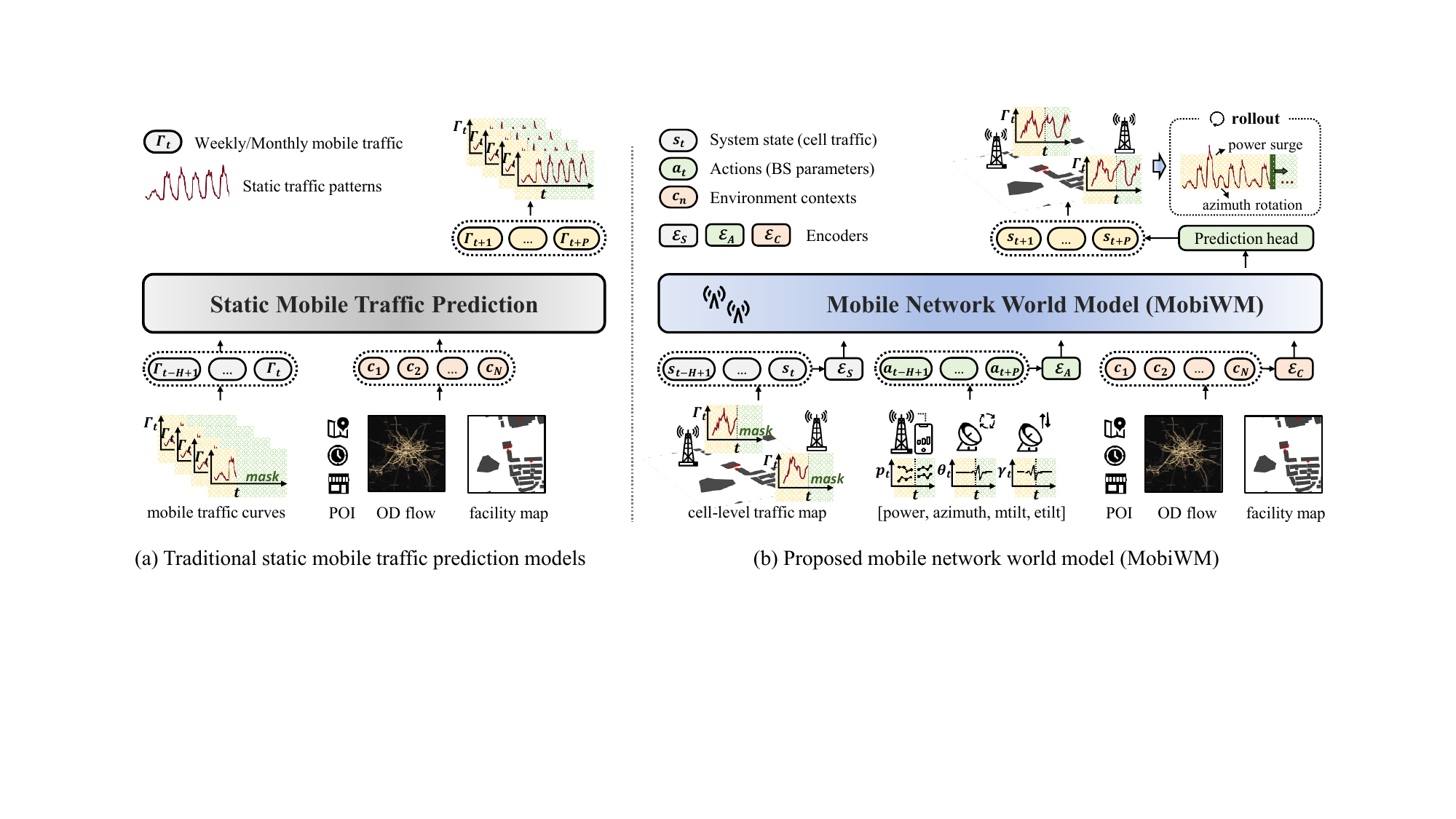}
\caption{Comparison of the traditional static mobile traffic prediction models and the proposed mobile network world model (MobiWM).}
\label{Fig_intro}
\end{figure*}

World models, learned simulators of environment dynamics for model-based reinforcement learning~\cite{ha2018world}, offer a natural framework to bridge this gap. By modelling how actions transform states over time, they learn action-state dynamics and enable forward rollout, counterfactual inference, and policy optimization in a latent space. STORM~\cite{zhang2023storm} and DreamerV3~\cite{hafner2023dreamerv3} demonstrate this paradigm through sample-efficient control and imagined trajectories across diverse tasks. World models have the capacity to capture how parameter adjustments drive state evolution, supporting counterfactual exploration and model-based network optimization.

In this paper, we propose a \textbf{Mobi}le network \textbf{W}orld \textbf{M}odel, \textbf{MobiWM}. Through an autoregressive world-modelling paradigm, MobiWM jointly learns the temporal and parameter dynamics of mobile networks. As shown in Figure~\ref{Fig_intro}, MobiWM treats cell-level traffic as states and network parameter adjustments as actions, and learns the transition from historical states and actions to future states. It conditions this transition on multimodal urban context, including Points of Interest (POI), Origin-Destination (OD) flows, and facility maps of buildings and BS layouts. An encoder-decoder backbone fuses states, actions, and context with shared spatial semantics, so that different modalities are aligned over the same geographic space. Factorized Spatio-Temporal Blocks (FSTBlocks) decouple topology and temporal modelling to capture non-local spatial effects and periodic temporal patterns. A graph-batch strategy with cell masking further supports maps with different numbers of cells, enabling map-level forecasting and iterative rollout over user-specified action trajectories. In summary, the main contributions are as follows:
\begin{itemize}
    \item We propose MobiWM, an action-conditioned autoregressive world model that jointly learns the temporal and parameter dynamics of mobile networks to support counterfactual network planning and optimization.
    \item We design FSTBlocks for factorized topology-temporal modelling and multimodal context encoders with learnable gating for fusing POI, OD, facility, and action information.
    \item We evaluate MobiWM on variable-parameter traffic data in Nanchang, China. MobiWM achieves strong rollout performance against traffic-prediction and world-model baselines, demonstrating its advantage in capturing network dynamics. A model-based Actor-critic case study further validates the potential of MobiWM for network planning and optimization.
\end{itemize}

\section{Related Work} 
\label{sec:rela}
\subsubsection{Mobile Traffic Prediction}
Mobile traffic prediction has evolved from statistical and traditional machine learning methods~\cite{shu2003arima_traffic,nikravesh2016sarima,du2020xgboost_bs} to deep spatio-temporal models~\cite{shi2015convlstm, zhang2017stresnet, yu2018stgcn, yang2024fedgtp, bettouche2025histm, wang2025mobimixer}. Recent work further incorporates environmental context and generative or foundation-model paradigms~\cite{zhang2026umask, mobifm2025}. However, most methods learn correlations from traffic observed under fixed or implicitly stationary network configurations. Parameter or topology changes alter coverage, association, and traffic redistribution, making these correlations unreliable and degrading prediction accuracy under interventions. Without explicitly modelling actions as causes of state transitions, these methods also cannot faithfully answer counterfactual ``what-if'' queries.

\medskip
\subsubsection{World Models}
World models provide a key learning mechanism for physical AI by modelling action-conditioned environment dynamics~\cite{ha2018world}. They represent states, actions, and context as high-dimensional vectors and learn their complex transition relations, capturing system evolution beyond static input--output correlations. These learned dynamics support iterative rollout, counterfactual reasoning, and model-based policy learning. Representative models, including DreamerV1--V3~\cite{hafner2020dreamerv1,hafner2021dreamerv2,hafner2023dreamerv3}, TransDreamer~\cite{chen2022transdreamer}, STORM~\cite{zhang2023storm}, and TD-MPC2~\cite{hansen2024tdmpc2}, demonstrate strong performance in control domains. Recent studies have also explored world models for wireless-network scheduling and edge intelligence~\cite{wang2025dmwm,zhao2025wm_edge,MobiWorld2025}. MobiWM extends this paradigm to cell-level traffic responses under continuous antenna-parameter trajectories, irregular network topology, and multimodal urban context.

\section{Preliminaries} 
\label{sec:pre}

\subsection{Mobile Traffic Dynamics}
\label{sec:dynamics}
Cellular traffic is determined by the aggregation of the actual service traffic generated by users within the coverage area of a cellular antenna. The distribution of user traffic is closely related to the urban environment where the cellular network is deployed. Different urban areas exhibit distinct numbers of users and user service behaviors, resulting in diverse spatiotemporal traffic distribution characteristics. The coverage area of a cell is influenced by antenna parameters, and users typically select the serving cell based on the reference signal received power (RSRP) from surrounding Base Stations (BSs). According to the 3GPP channel model~\cite{3gpp_38901}, the RSRP at location $\mathbf{r}$ from cell $v_i$ can be denoted as
\begin{equation}
    P_{\mathrm{rx}}^{i}(\mathbf{r})\text{(dB)} = p_t^i + G^{i}(\theta_t^i, \gamma_{\mathrm{m},t}^i, \gamma_{\mathrm{e},t}^i) - \mathrm{PL}(\mathbf{r}, f_c) + u,
\end{equation}
where $p_t^i$ denotes the transmit power of cell $v_i$. $G^{i}$ represents the antenna gain, which is characterized by the azimuth angle $\theta_t^i$, as well as the mechanical and electrical downtilts $\gamma_{\mathrm{m},t}^i$ and $\gamma_{\mathrm{e},t}^i$. $\mathrm{PL}(\cdot)$ denotes the path loss, which is a function of the Tx-Rx distance and the carrier frequency $f_c$, where $f_c$ is typically fixed in practical deployments. The term $u$ represents the residual component associated with small-scale fading effects, such as multipath propagation.

Therefore, the coverage region of a cell, denoted as $\mathcal{R}$, can be formulated as a function of $p_t^i$, $\theta_t^i$, $\gamma_{\mathrm{m},t}^i$, $\gamma_{\mathrm{e},t}^i$, and the urban environment $\mathbf{c}$. Meanwhile, user service traffic can be abstracted as a traffic density $\rho_t(\mathbf{r})$, which is influenced by the urban environment $\mathbf{c}$. Accordingly, mobile traffic dynamics can be characterized by two coupled processes: the \textit{parameter dynamics} that affect the evolution of $\mathcal{R}$ and the \textit{spatiotemporal dynamics} that govern the variation of $\rho_t(\mathbf{r})$. The governing formulation can be expressed as:
\begin{equation}
\label{eq:traffic}
    s_t^i = \int_{\mathbf{r} \in \mathcal{R}} \mathbf{1}\Big[i = \arg\max_{j \in \mathcal{V}} P_{\mathrm{rx}}^{j}(\mathbf{r})\Big]\, \rho_t(\mathbf{r})\, \mathrm{d}\mathbf{r},
\end{equation}
where $\mathcal{V}=\{v_1,\dots,v_N\}$ is the set of cells in the map.

\subsection{Problem Formulation}
\subsubsection{System Representation}
We model a cellular network with $N$ cells as a directed graph $\mathbf{G}=(\mathcal{V},\mathcal{E})$, where $\mathcal{V}$ and edges $\mathcal{E}$ encode spatial adjacency. At time $t < T$, the state $\mathbf{s}_t=[s_t^1,\dots,s_t^N]\in\mathbb{R}^{N}$ records cell traffic states, the action is defined as $\mathbf{a}_t=[p_t^i, \theta_t^i, \gamma_{\mathrm{m},t}^i, \gamma_{\mathrm{e},t}^i]\in\mathbb{R}^{N\times4}$. The urban context $\mathbf{c}=\{\mathbf{c}^{\mathrm{poi}}\in\mathbb{R}^{S\times S\times K},\mathbf{c}^{\mathrm{od}}\in\mathbb{R}^{S\times S\times T},\mathbf{c}^{\mathrm{fac}}\in\mathbb{R}^{H_f\times W_f\times 1}\}$ represents POI, OD flow, and facility information, where $S$ denotes coarse urban grids, and $H_f$/$W_f$ represents fine pixel grids. $K$ is the dimension of POIs.

\medskip
\subsubsection{Mobile Network World modelling}
The objective is to learn a parameterized dynamics model $f_{\Omega}$ that predicts future states from historical states, historical and planned actions, and environmental context:
\begin{equation}
\label{eq:wm}
    \hat{\mathbf{s}}_{t+1:t+P} = f_{\Omega}\big(\mathbf{s}_{t-H+1:t},\; \mathbf{a}_{t-H+1:t+P},\; \mathbf{c}\big),
\end{equation}
where $H$ is the historical window length, $P$ is the prediction horizon, and $\Omega$ denotes the learnable parameters. The planned-action segment $\mathbf{a}_{t+1:t+P}$ specifies the parameter trajectory for the prediction interval. For a longer trajectory, predicted states can be fed back with the actions for the next segment. Let $\tilde{\mathbf{s}}_\tau=\mathbf{s}_\tau$ for observed steps $\tau\leq t$ and $\tilde{\mathbf{s}}_\tau=\hat{\mathbf{s}}_\tau$ for predicted steps $\tau>t$. The $k$-th rollout segment is computed as
\begin{equation}
\label{eq:rollout}
    \begin{aligned}
    \hat{\mathbf{s}}_{t+kP+1:t+(k+1)P}
    &= f_{\Omega}\big(
    \tilde{\mathbf{s}}_{t+kP-H+1:t+kP},\\
    &\qquad \mathbf{a}_{t+kP-H+1:t+(k+1)P}, \mathbf{c}\big),
    \end{aligned}
\end{equation}
Repeating this procedure enables multi-segment counterfactual rollout over a specified action sequence, while errors may accumulate as the rollout horizon grows.

\section{Methods} 
\label{sec:method}
\begin{figure}[t]
    \centering
    \includegraphics[width=\linewidth]{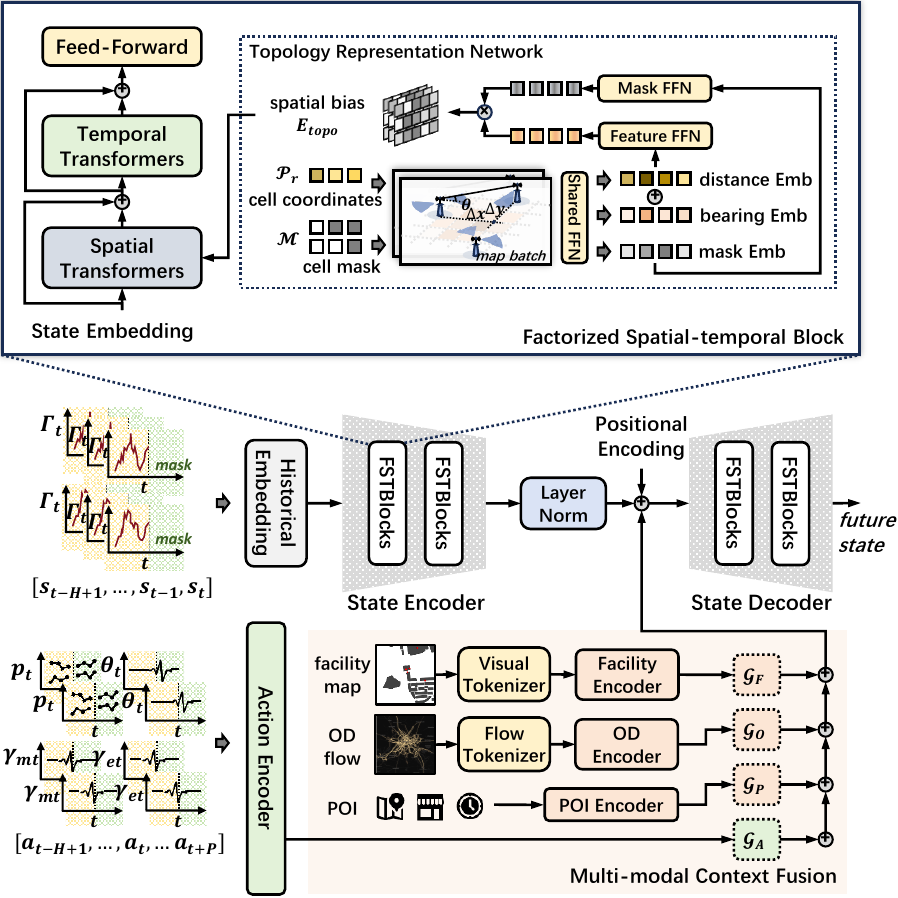}
    \caption{Overview of MobiWM. It is composed of an encoder-decoder backbone with factorized spatio-temporal blocks (FSTBlocks) for dynamics learning, conditioned on actions and multimodal environmental context.}
    \label{fig:overview}
\end{figure}

We propose MobiWM, a world model for mobile networks that learns the dynamics between network parameter adjustments and traffic variations through an encoder-decoder architecture. The model is designed to capture the complex spatio-temporal dependencies and topological features of mobile networks while fusing multimodal environmental context. Figure~\ref{fig:overview} illustrates the overall architecture of MobiWM.

\subsection{Base Model}
The base model is an encoder-decoder architecture for state dynamics learning, combined with graph batching and cell masking for map-level forecasting.

\subsubsection{Graph Batch for Irregular Network Topology}
Cells are irregularly deployed due to terrain, population density, and infrastructure constraints. We organize all $N$ cells in a district-level map as one graph $\mathscr g=(\mathscr v,\mathscr e)$, where each node carries its state and action series. This differs from conventional traffic forecasting that treats each station independently: the whole map is predicted as one sample so that inter-cell effects remain visible to the model. Maps with different cell counts are zero-padded for batching~\cite{hamilton2017graphsage, hu2020ogb}.

\subsubsection{Encoder-Decoder State Dynamics Model}
MobiWM utilizes a state encoder and a state decoder to learn the traffic dynamics and predict the future states. The historical state $[\mathbf{s}_{t-H+1}, \dots, \mathbf{s}_t]$ is projected to a $d_{\mathrm{enc}}$-dimensional token sequence $\mathbf{S}^{0}\in\mathbb{R}^{N\times H\times d_{\mathrm{enc}}}$ by an embedding layer. Each token corresponds to one cell at one time step, allowing the model to preserve both spatial and temporal axes. A state encoder $\mathcal{E}_S$ with $L_e$ FSTBlocks compresses the history into
\begin{equation}
\label{eq:encoder}
    \mathbf{Z}_H = \mathcal{E}_S\big(\mathbf{S}^{0}\big) \in \mathbb{R}^{N \times H \times d_{\mathrm{enc}}}.
\end{equation}
The decoder $\mathcal{D}_S$ with $L_d$ FSTBlocks operates on the condition-augmented encoder latent $\mathbf{Z}_0$ defined in Eq.~\eqref{eq:fusion}:
\begin{equation}
\label{eq:decoder}
    \mathbf{Z}_P = \mathcal{D}_S\big(\mathbf{Z}_0\big) \in \mathbb{R}^{N \times P \times d_{\mathrm{dec}}}.
\end{equation}
Finally, the decoder output is mapped back to traffic values by a linear prediction head:
\begin{equation}
\label{eq:output}
    \hat{\mathbf{s}}_{t+1:t+P} = \mathcal{O}_\text{out}\big(\mathbf{Z}_P).
\end{equation}

\subsubsection{Action Encoding}
The action encoder $\mathcal{E}_A$ embeds historical and future actions into the decoder latent space:
\begin{equation}
\label{eq:action_enc}
    \mathbf{h}_A = \mathcal{E}_A\big(\mathbf{a}_{t-H+1:t+P}\big) \in \mathbb{R}^{N \times (H+P) \times d_{\mathrm{dec}}}.
\end{equation}
Future actions allow the decoder to condition on hypothetical parameter changes that have not yet occurred in the observed history. During rollout (Eq.~\eqref{eq:rollout}), operators can specify future action trajectories for counterfactual analysis.

\subsection{Factorized Spatio-Temporal Block}
\label{sec:fstblock}
Mobile traffic has spatial regularity from network topology and temporal regularity from daily and weekly activity. Directly applying joint spatio-temporal attention over $N$ cells and $T$ steps costs $\mathcal{O}(N^2T^2)$, which is impractical for large maps. MobiWM therefore factorizes it into spatial and temporal stages with complexity $\mathcal{O}(N^2T+NT^2)$, while still allowing information to propagate across both axes.

\subsubsection{Factorized Attentions}
Each FSTBlock applies spatial self-attention over cell tokens at each time step, temporal self-attention over each cell's time series, and a position-wise FFN with residual connections. The spatial stage injects a topology-aware bias $\mathbf{E}_{\mathrm{topo}}$ to encode geometric relations between cells before temporal modelling. The temporal stage then captures periodicity and long-range trends on spatially enriched tokens. 

\subsubsection{Topology-based Spatial Bias}
\label{sec:topo_bias}
To make spatial attention topology-aware, a Topology Representation Network (TRN) computes a pairwise bias from cell geometry and the cell mask. Given cell coordinates $\mathcal{P}_r=\{(x_i,y_i)\}_{i=1}^{N}$, we derive relative displacement, distance $d_{ij}$, and bearing angle $\alpha_{ij}$ for each cell pair. These features encode both proximity and orientation, which are important because antenna direction and cell layout affect interference and traffic redistribution. They are projected to $\mathbf{e}_{ij}^{\mathrm{feat}}$ and combined with a mask embedding $\mathbf{e}_{ij}^{\mathrm{mask}}$:
\begin{equation}
\label{eq:topo_bias}
    \mathbf{E}_{\mathrm{topo}}[i,j] = \mathbf{e}_{ij}^{\mathrm{feat}} \otimes \mathbf{e}_{ij}^{\mathrm{mask}} \in \mathbb{R}^{n_h},
\end{equation}
where $n_h$ is the number of attention heads. The bias is added to the spatial attention logits:
\begin{equation}
\label{eq:biased_attn}
    \mathrm{SpatialAttn}(\mathbf{X}) = \mathrm{softmax}\left(\frac{\mathbf{Q}\mathbf{K}^\top}{\sqrt{d_k}} + \mathbf{E}_{\mathrm{topo}}\right)\mathbf{V},
\end{equation}
where $\mathbf{Q}, \mathbf{K}, \mathbf{V}$ are the query, key, and value projections of $\mathbf{X}$, and $d_k$ is the key dimension. 

\subsection{Environmental Context Fusion}
\label{sec:conditioning}
MobiWM fuses environmental context through modality-specific encoding, shared positional encoding, and learnable gating. Each modality has a different data structure; thus, their spatial semantics must be aligned before being used by the decoder.

\subsubsection{Multi-modal Context Encoding}
MobiWM uses three encoders to encode POI, OD flow, and facility maps with different modalities into latent embeddings within a shared condition space. The context encoding is denoted as:
\begin{equation}
\begin{aligned}
    \mathbf{h}_P &= \mathrm{POIEnc}(\mathbf{c}^{\mathrm{poi}}),\\
    \mathbf{h}_O &= \mathrm{ODEnc}(\mathbf{c}^{\mathrm{od}}), \\
    (\mathbf{h}_{F_f}, \mathbf{h}_{F_c}, \mathbf{h}_{F}) &= \mathrm{FacEnc}(\mathbf{c}^{\mathrm{fac}}), \\
\end{aligned}
\label{eq:cond_enc}
\end{equation}
where $\mathbf{h}_P,~\mathbf{h}_O,~\mathbf{h}_{F_f},~\mathbf{h}_{F_c},~\mathbf{h}_{F}\in \mathbb{R}^{d_{\mathrm{dec}}}$. $\mathbf{h}_{F_f}$ and $\mathbf{h}_{F_c}$ are fine and coarse embeddings of facility maps designed for multi-granularity context fusion. 

\subsubsection{Multi-granularity Positional Encoding}
\label{sec:pe}
Temporal encoding maps the time-slot index $q_t$ and day-of-week index $w_t$ to learnable embeddings:
\begin{equation}
\label{eq:temporal_pe}
    \mathbf{p}_t^{\mathrm{temp}} = \mathbf{E}_{\mathrm{slot}}(q_t) + \mathbf{E}_{\mathrm{dow}}(w_t) \in \mathbb{R}^{d_{\mathrm{dec}}},
\end{equation}
where $\mathbf{E}_{\mathrm{slot}}\in\mathbb{R}^{Q\times d_{\mathrm{dec}}}$ and $\mathbf{E}_{\mathrm{dow}}\in\mathbb{R}^{7\times d_{\mathrm{dec}}}$ encode diurnal and weekly cycles. For spatial encoding, a key challenge is that cell-level states, grid-level POI/OD, and pixel-level facility maps share the same geographic space but differ in granularity. We design a shared Fourier-based positional encoding: for any coordinate $\mathbf{r} = (x, y)$, we normalize it and project through $L$ log-spaced frequency bands to obtain Fourier features, then map them via a single shared-parameter MLP:
\begin{equation}
\label{eq:fac_pe1}
    \mathbf{p}^{\mathrm{spat}}(\mathbf{r}) = \mathrm{MLP}_{\mathrm{shared}}\Big(\big[\sin(\omega_l \bar{\mathbf{r}}),\; \cos(\omega_l \bar{\mathbf{r}})\big]_{l=1}^{L}\Big) \in \mathbb{R}^{d_{\mathrm{dec}}},
\end{equation}
where $\bar{\mathbf{r}}$ is the coordinate normalized by the maximum spatial extent and $\{\omega_l\}_{l=1}^{L}$ are $L$ logarithmically spaced frequency bands. $\mathrm{MLP}_{\mathrm{shared}}$ is then applied to cell coordinates, $S{\times}S$ coarse grid centers, and $H_f{\times}W_f$ fine pixel grids. Let $\mathbf{p}^{\mathrm{spat}}_{\mathrm{fine}}$ and $\mathbf{p}^{\mathrm{spat}}_{\mathrm{coarse}}$ denote encodings on fine and coarse grids. They are injected into intermediate feature maps of each context encoder before aggregation:
\begin{equation}
\label{eq:fac_pe2}
    \mathbf{h}_{F_f} \leftarrow \mathbf{h}_{F_f} + \mathbf{p}^{\mathrm{spat}}_{\mathrm{fine}},\quad \mathbf{h}_{F_c} \leftarrow \mathbf{h}_{F_c} + \mathbf{p}^{\mathrm{spat}}_{\mathrm{coarse}}.
\end{equation}
$\mathbf{h}_P$ and $\mathbf{h}_O$ similarly receive $\mathbf{p}^{\mathrm{spat}}_{\mathrm{coarse}}$. Thus, tokens from different granularities but similar locations receive compatible positional signals. We denote the resulting context set as $\mathbf{h}_C=\{\mathbf{h}_P,\mathbf{h}_O,\mathbf{h}_{F}+\mathbf{h}_{F_f}+\mathbf{h}_{F_c}\}$.

\subsubsection{Learnable Gating}
\label{sec:gating}
Context and action embeddings are integrated into the encoded state through four gates $\mathcal{G}_F,\mathcal{G}_O,\mathcal{G}_P,\mathcal{G}_A$. The gates learn global modality-specific importance weights:
\begin{equation}
\label{eq:gating}
    \mathcal{G}_i(\bar{\mathbf{h}}_i) = \sigma(g_i) \cdot \phi_i(\bar{\mathbf{h}}_i), ~\bar{\mathbf{h}}_i=\mathcal{B}_i(\mathbf{h}_i)\in\mathbb{R}^{N\times P\times d_{\mathrm{dec}}},
\end{equation}
where $\mathcal{B}_A$ performs temporal alignment to the future steps. For POI, OD, and facility features, $\mathcal{B}_i$ uses the shared spatial encodings to aggregate features at cell locations and broadcasts static features over the $P$ decoder steps; time-varying OD features retain their corresponding temporal indices. Each aligned condition $\bar{\mathbf{h}}_i$ is projected by a linear layer $\phi_i(\cdot)$, and a learnable scalar gate $g_i$ controls its contribution. After that, the shape-aligned gated representations are then added to the encoded state:
\begin{equation}
\label{eq:fusion}
    \mathbf{Z}_H\leftarrow \mathbf{Z}_H \oplus \sum_{i\in\{F,O,P,A\}}\mathcal{G}_i(\bar{\mathbf{h}}_i),
\end{equation}
where $\oplus$ means combining after mapping. These scalar gates provide a way to learn modality importance during training.

\section{Experiments} 
\label{sec:eval}

\subsection{Experimental Setup}

\subsubsection{Dataset}
\label{sec:dataset}
Operator logs are commercially sensitive and perturbing live BS configurations is costly and service-disruptive; real variable-parameter traffic data cannot be collected at the required scale. We construct a simulation-augmented dataset covering 31,900 cells across 9 districts of Nanchang, China, at 15-minute granularity over one week ($T{=}672$). This dataset can be used as reliable training data for world models: traffic demand and its temporal variation, BS deployment and initial parameters, OD mobility, POIs, and urban geometry come from real network or geographic data; Simulations are based on the Sionna platform and comply with existing 5G protocol stack standards.

For constructing the dataset, each district is divided into $300\,\mathrm{m}\times300\,\mathrm{m}$ maps, and we use OpenStreetMap\footnote{\url{https://www.openstreetmap.org/}} buildings to form the 3D scenes, and UE locations are sampled from WorldPop\footnote{\url{https://www.worldpop.org/}}. We disaggregate measured cell traffic to UE demand while preserving its temporal profile, sample parameter trajectories, and use Sionna~\cite{hoydis2023sionna} to recompute RSRP. UE associations are updated by 3GPP initial-access and A3 handover rules~\cite{3gpp_38304,3gpp_38331}, after which the unchanged UE demand is reaggregated over the new associations. The resulting \emph{Para} and \emph{Topo} subsets with \emph{Urban} and \emph{Suburb} regions model parameter changes under fixed topology and cell activation/deactivation, respectively.

\subsubsection{Model Configuration and Training Details}
We use MobiWM-M for the main comparison and report all four scales in the efficiency study. Model scale (S/M/L/XL) is a swept hyperparameter. The 90\%/10\% split is performed by spatial grid: each grid belongs exclusively to either training or testing, ensuring no geographic overlap between the two sets. Multi-segment rollout in Eq.~\eqref{eq:rollout} is used only for inference and evaluation. Table~\ref{tab:model_training_config} summarizes the implementation details, where each scale is represented by $(L_{\mathrm{enc}},L_{\mathrm{dec}},d_{\mathrm{enc}},d_{\mathrm{dec}},n_h)$, denoting the numbers of encoder/decoder FSTBlocks, encoder/decoder dimensions, and attention heads. Models are trained and tested on a server equipped with a single NVIDIA RTX 4090 GPU and two Intel Xeon Platinum 8358 CPUs at 2.60 GHz.
\begin{table}[t]
\centering
\caption{MobiWM configurations and training settings.}
\label{tab:model_training_config}
\fontsize{8pt}{9pt}\selectfont
\setlength{\tabcolsep}{1pt}
\renewcommand{\arraystretch}{1.05}
\begin{tabular}{@{}C{0.20\columnwidth}C{0.27\columnwidth}C{0.20\columnwidth}C{0.27\columnwidth}@{}}
\toprule
\textbf{Parameter} & \textbf{Value} & \textbf{Parameter} & \textbf{Value} \\
\midrule
MobiWM-S & $(2,2,64,64,4)$ & MobiWM-M & $(4,3,96,96,6)$ \\
MobiWM-L & $(4,4,128,96,8)$ & MobiWM-XL & $(6,6,128,128,8)$ \\
POI Dim.($K$) & 21 & His./Horizon & 16 / 4 \\
Window stride & 4 & Rollout length & 672 \\
Epochs/warmup & 200 / 5 & LR schedule & Linear + cosine \\
Batch size & 256 & Num. of runs & 8 \\
Train/test split & 90\% / 10\% & Random seed & 42 \\
Power range & [0, 80] & Azimuth range & [0, 360] \\
Mtilt range & [0, 15] & Etilt range & [0, 15] \\
\bottomrule
\end{tabular}
\end{table}

\subsubsection{Baselines}
We compare MobiWM against three categories of baselines.
(i)~Mobile traffic prediction models: FedGTP~\cite{yang2024fedgtp}, HiSTM~\cite{bettouche2025histm}, and MobiFM~\cite{mobifm2025}. For each, we evaluate both the original model (which predicts without action conditioning) and a world-model variant (suffixed with \textbf{-WM}) that augments the original architecture with our action-state formulation.
(ii)~Spatio-temporal prediction models: iTransformer~\cite{liu2024itransformer}, Informer~\cite{zhou2021informer}, TimeMoE~\cite{shi2024timemoe}, and CSDI~\cite{tashiro2021csdi}, all adapted to the world-model formulation to accept action inputs.
(iii)~Representative world models: TD-MPC2~\cite{hansen2024tdmpc2}, STORM~\cite{zhang2023storm}, and DreamerV3~\cite{hafner2023dreamerv3}, which natively support action-conditioned state prediction.

\subsubsection{Evaluation Metrics} We evaluate the rollout performance of MobiWM and baselines using three metrics: Jensen-Shannon Divergence (JSD) to measure distributional similarity between predicted and true traffic distributions, Mean Absolute Error (MAE) to quantify the average magnitude of prediction errors, and Normalized Root Mean Square Error (NRMSE) to assess the overall prediction accuracy normalized by the range of true values. These metrics together provide a comprehensive evaluation of both the fidelity and accuracy.

\subsection{Overall Rollout Performance}
\begin{table*}[t]
\centering
\fontsize{7.5pt}{8.5pt}\selectfont
\newcommand{\resultci}[2]{%
    #1%
    {\fontsize{1pt}{1pt}\selectfont$\pm$}%
    {\fontsize{4.5pt}{4.5pt}\selectfont#2}%
  }
\caption{Performance comparison across different scenarios. Small values to the right of the means are 95\% confidence half-widths. $\bar{\Delta}$ denotes the average percentage improvement. (\colorbox{orange!25}{\textbf{Bold}} indicates best, \colorbox{cyan!20}{\underline{underlined}} indicates second best.)}
\label{tab:performance_comparison}
\setlength{\tabcolsep}{1pt} 
\resizebox{\textwidth}{!}{%
\begin{tabular}{c|ccc|ccc|ccc|ccc|c}
\toprule
\multirow{3.5}{*}{\textbf{Model}} & \multicolumn{3}{c|}{\textbf{Urban-Para}} & \multicolumn{3}{c|}{\textbf{Urban-Topo}} & \multicolumn{3}{c|}{\textbf{Suburb-Para}} & \multicolumn{3}{c|}{\textbf{Suburb-Topo}} & \multirow{3.5}{*}{$\bar{\Delta}$} \\
\cmidrule(lr){2-4} \cmidrule(lr){5-7} \cmidrule(lr){8-10} \cmidrule(lr){11-13}
& \tabincell{c}{JSD} & \tabincell{c}{MAE\\($\times 1\text{e}5$)} & \tabincell{c}{NRMSE}
& \tabincell{c}{JSD} & \tabincell{c}{MAE\\($\times 1\text{e}5$)} & \tabincell{c}{NRMSE}
& \tabincell{c}{JSD} & \tabincell{c}{MAE\\($\times 1\text{e}5$)} & \tabincell{c}{NRMSE}
& \tabincell{c}{JSD} & \tabincell{c}{MAE\\($\times 1\text{e}5$)} & \tabincell{c}{NRMSE} & \\
\midrule
FedGTP & \resultci{0.783}{.00350} & \resultci{3.99}{.440} & \resultci{0.782}{.0789} & \resultci{0.764}{.00620} & \resultci{3.32}{.283} & \resultci{0.876}{.138} & \resultci{0.763}{.00750} & \resultci{3.79}{.474} & \resultci{0.729}{.0607} & \resultci{0.790}{.00200} & \resultci{2.98}{.279} & \resultci{0.897}{.137} & 29.0\% \\
FedGTP-WM & \resultci{0.474}{.0281} & \resultci{3.38}{.433} & \resultci{0.752}{.0638} & \resultci{0.556}{.0168} & \resultci{2.96}{.247} & \resultci{0.898}{.133} & \resultci{0.455}{.0319} & \resultci{3.67}{.497} & \resultci{0.723}{.0628} & \resultci{0.515}{.0216} & \resultci{2.76}{.343} & \resultci{0.843}{.103} & 17.6\% \\
HiSTM & \resultci{0.487}{.0267} & \resultci{4.11}{.439} & \resultci{0.850}{.122} & \resultci{0.537}{.0281} & \resultci{3.92}{.424} & \resultci{1.15}{.294} & \resultci{0.515}{.0319} & \resultci{5.20}{.667} & \resultci{1.13}{.292} & \resultci{0.556}{.0296} & \resultci{5.17}{.964} & \resultci{1.89}{.766} & 36.8\% \\
HiSTM-WM & \resultci{0.506}{.0288} & \resultci{3.93}{.492} & \resultci{0.775}{.0890} & \resultci{0.491}{.0273} & \resultci{3.40}{.360} & \resultci{0.942}{.170} & \resultci{0.439}{.0316} & \resultci{3.80}{.496} & \resultci{0.772}{.125} & \resultci{0.566}{.0237} & \resultci{2.92}{.357} & \resultci{0.886}{.135} & 21.9\% \\
MobiFM & \cellcolor{cyan!20}\resultci{\underline{0.341}}{.0456} & \resultci{4.74}{.742} & \resultci{1.08}{.204} & \resultci{0.485}{.0337} & \resultci{6.41}{.537} & \resultci{2.12}{.602} & \resultci{0.423}{.0339} & \resultci{4.28}{.725} & \resultci{0.853}{.104} & \resultci{0.510}{.0373} & \resultci{2.94}{.672} & \resultci{0.863}{.0392} & 30.0\% \\
MobiFM-WM & \resultci{0.344}{.0270} & \cellcolor{orange!25}\resultci{\textbf{2.97}}{.305} & \cellcolor{orange!25}\resultci{\textbf{0.663}}{.0783} & \cellcolor{cyan!20}\resultci{\underline{0.422}}{.0309} & \cellcolor{cyan!20}\resultci{\underline{2.90}}{.438} & \resultci{0.872}{.208} & \resultci{0.378}{.0300} & \resultci{3.78}{.437} & \resultci{0.813}{.119} & \resultci{0.440}{.0409} & \resultci{4.48}{.800} & \resultci{1.53}{.409} & \cellcolor{cyan!20}\underline{16.4\%} \\
\midrule
iTransformer-WM & \resultci{0.478}{.0317} & \resultci{3.33}{.555} & \resultci{0.739}{.0555} & \resultci{0.622}{.0439} & \resultci{3.48}{.292} & \resultci{1.94}{.583} & \resultci{0.539}{.0417} & \resultci{4.72}{.792} & \resultci{0.877}{.0909} & \resultci{0.641}{.0308} & \resultci{4.68}{.414} & \resultci{1.51}{.287} & 34.3\% \\
Informer-WM & \resultci{0.467}{.0371} & \resultci{3.74}{.622} & \resultci{0.814}{.101} & \resultci{0.609}{.0432} & \resultci{3.72}{.261} & \resultci{0.973}{.270} & \resultci{0.471}{.0268} & \resultci{3.81}{.645} & \resultci{0.714}{.0354} & \resultci{0.651}{.0275} & \resultci{5.10}{.442} & \resultci{1.64}{.305} & 29.9\% \\
TimeMoE-WM & \resultci{0.430}{.0309} & \resultci{3.35}{.491} & \resultci{0.696}{.0373} & \resultci{0.523}{.0273} & \resultci{4.79}{.735} & \resultci{1.31}{.373} & \resultci{0.488}{.0268} & \resultci{3.49}{.469} & \cellcolor{cyan!20}\resultci{\underline{0.707}}{.0561} & \resultci{0.408}{.0363} & \resultci{2.96}{.540} & \resultci{1.11}{.503} & 22.1\% \\
CSDI-WM & \resultci{0.520}{.0487} & \resultci{6.04}{.399} & \resultci{2.11}{.301} & \resultci{0.432}{.0361} & \resultci{12.1}{.505} & \resultci{4.66}{.912} & \resultci{0.498}{.0336} & \resultci{9.35}{.837} & \resultci{2.60}{.0772} & \resultci{0.515}{.0435} & \resultci{15.4}{.596} & \resultci{6.96}{.480} & 59.8\% \\
\midrule
TD-MPC2 & \resultci{0.639}{.0445} & \resultci{4.09}{.542} & \resultci{0.966}{.169} & \resultci{0.667}{.0144} & \resultci{3.07}{.349} & \cellcolor{cyan!20}\resultci{\underline{0.796}}{.0931} & \resultci{0.711}{.0146} & \resultci{3.82}{.473} & \resultci{0.935}{.0924} & \resultci{0.613}{.0455} & \resultci{5.30}{1.13} & \resultci{1.61}{.505} & 39.3\% \\
STORM & \resultci{0.634}{.0279} & \resultci{3.94}{.499} & \resultci{0.758}{.0566} & \resultci{0.629}{.0204} & \resultci{3.09}{.370} & \resultci{0.799}{.0925} & \resultci{0.703}{.0161} & \resultci{4.11}{.474} & \resultci{0.784}{.0963} & \resultci{0.666}{.0212} & \resultci{2.78}{.363} & \resultci{0.806}{.0696} & 24.7\% \\
DreamerV3 & \resultci{0.617}{.0343} & \resultci{3.79}{.585} & \resultci{0.765}{.0357} & \resultci{0.545}{.0245} & \resultci{3.08}{.362} & \resultci{0.825}{.147} & \resultci{0.541}{.0277} & \resultci{3.62}{.503} & \resultci{0.722}{.0426} & \resultci{0.538}{.0308} & \cellcolor{cyan!20}\resultci{\underline{2.30}}{.331} & \cellcolor{orange!25}\resultci{\textbf{0.727}}{.0749} & 18.0\% \\
\midrule
\textbf{MobiWM (Ours)} & \cellcolor{orange!25}\resultci{\textbf{0.334}}{.0330} & \cellcolor{cyan!20}\resultci{\underline{3.06}}{.363} & \cellcolor{cyan!20}\resultci{\underline{0.673}}{.0892} & \cellcolor{orange!25}\resultci{\textbf{0.324}}{.0219} & \cellcolor{orange!25}\resultci{\textbf{2.89}}{.365} & \cellcolor{orange!25}\resultci{\textbf{0.793}}{.171} & \cellcolor{orange!25}\resultci{\textbf{0.311}}{.0295} & \cellcolor{orange!25}\resultci{\textbf{3.14}}{.330} & \cellcolor{orange!25}\resultci{\textbf{0.688}}{.166} & \cellcolor{orange!25}\resultci{\textbf{0.369}}{.0260} & \cellcolor{orange!25}\resultci{\textbf{2.15}}{.299} & \cellcolor{cyan!20}\resultci{\underline{0.799}}{.156} & - \\
\bottomrule
\end{tabular}%
}
\end{table*}

\begin{figure}[t]
    \centering
    \includegraphics[width=\linewidth]{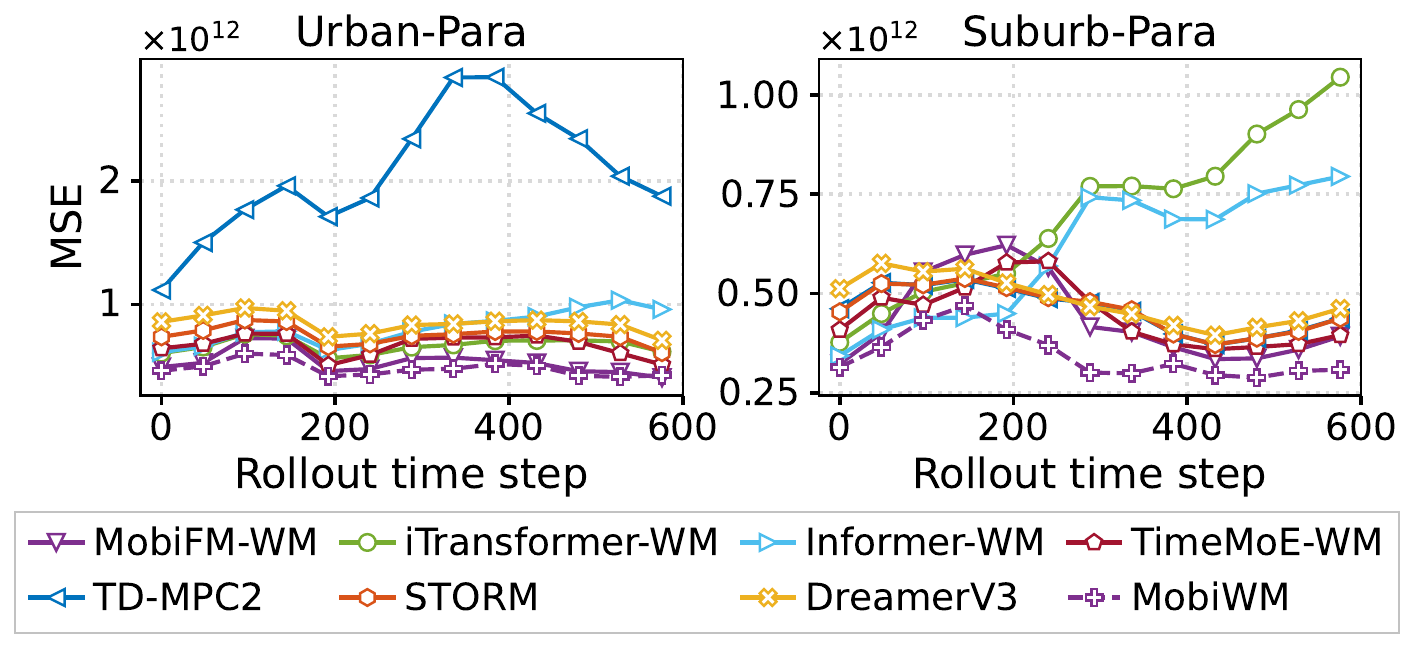}
    \caption{Rollout error temporal variation curve.}
    \label{fig:rollout}
\end{figure}

Table~\ref{tab:performance_comparison} reports one-week rollouts across \emph{Urban-Para}, \emph{Urban-Topo}, \emph{Suburb-Para}, and \emph{Suburb-Topo} datasets. The performance differences are confirmed to be statistically significant through two-sided Wilcoxon signed-rank tests with Holm correction. MobiWM achieves the best overall performance in all four scenarios, demonstrating strong distributional fidelity and competitive pointwise accuracy. Compared with action-conditioned temporal predictors (iTransformer-WM, Informer-WM, TimeMoE-WM, and CSDI-WM), MobiWM yields average improvements of 22.11-59.79\%. These predictors learn a one-shot conditional mapping to a future sequence, whereas MobiWM explicitly learns action-conditioned state transitions and composes them autoregressively, better preserving temporal regularities and action effects over long horizons. Likewise, the WM variants of FedGTP, HiSTM, and MobiFM generally outperform their original versions, demonstrating the benefit of explicitly modelling parameter dynamics. Figure~\ref{fig:rollout} further shows the temporal error distribution over the one-week rollout. MobiWM maintains the lowest and most stable error, corroborating the advantage of autoregressively composing learned action-state transitions.

\subsection{Ablation Studies}
\begin{figure}[t]
    \centering
    \includegraphics[width=0.9\linewidth]{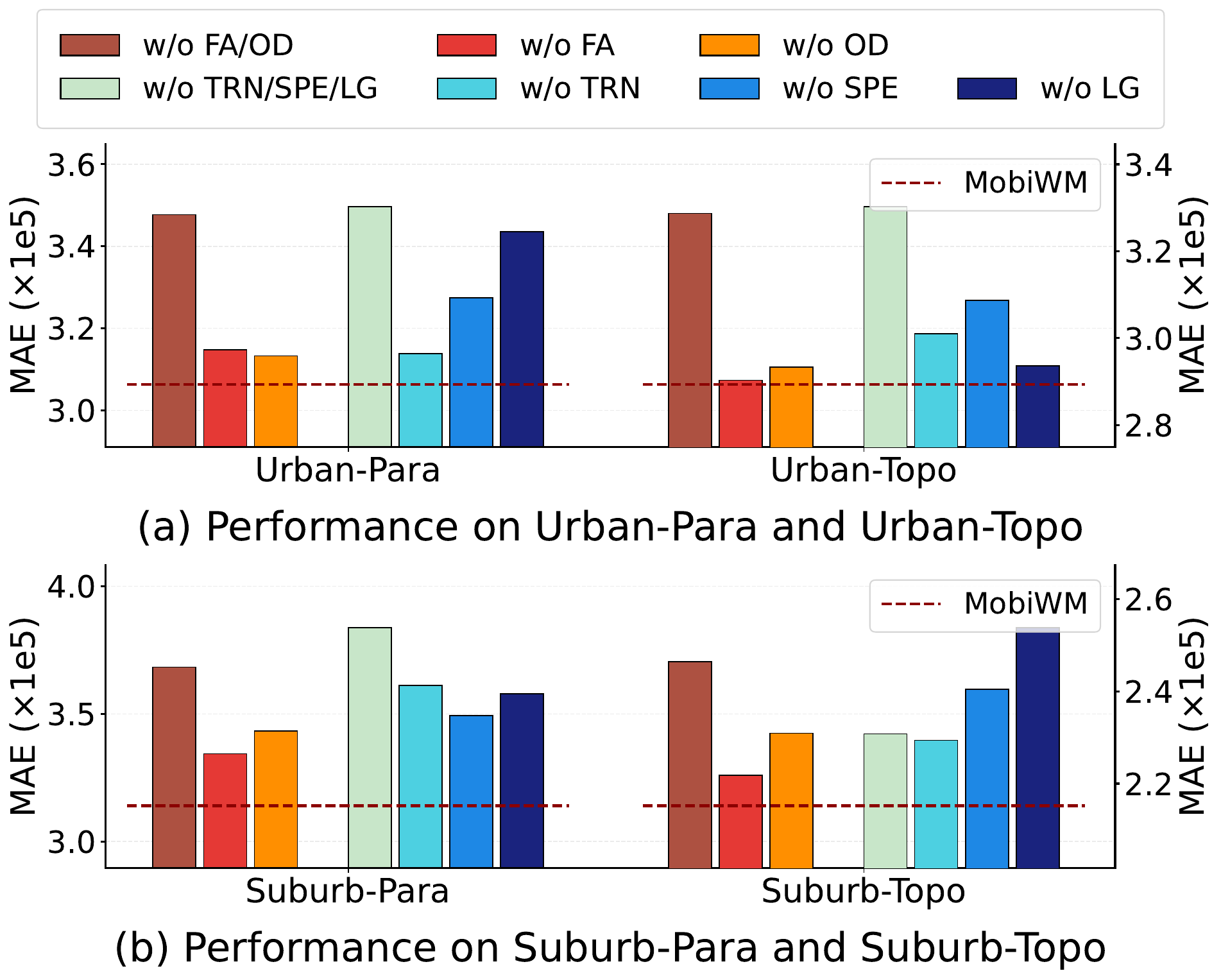}
    \caption{Ablation study on environment context modalities (w/o FA, w/o OD) and fusion mechanism components (w/o TRN, w/o SPE, w/o LG). The dashed line marks the full MobiWM.}
    \label{fig:ablation}
\end{figure}

\subsubsection{Environment context ablation}
As reported in Figure~\ref{fig:ablation}, removing the facility map (w/o FA) causes consistent MAE increases across all four scenarios, showing that static infrastructure priors and time-varying mobility demand are both important. The facility map helps characterize coverage constraints induced by buildings and BS layouts, while OD flow reflects human mobility that drives traffic redistribution. Removing both causes the largest context-related degradation, confirming their complementary roles in accurate rollout.

\subsubsection{Multimodal fusion ablation}
As reported in Figure~\ref{fig:ablation}, removing all fusion components (w/o TRN/SPE/LG) leads to the largest performance drop, indicating that fusion design is as important as modality selection. Among individual components, shared positional encoding and learnable gating contribute strongly by aligning spatial semantics and adapting modality weights, while TRN provides additional topology-aware bias for irregular cell relations. The full design achieves the lowest MAE, validating their synergy.

\subsection{Discussion of Dynamics Learning}
\begin{figure}[t]
    \centering
    \includegraphics[width=\linewidth]{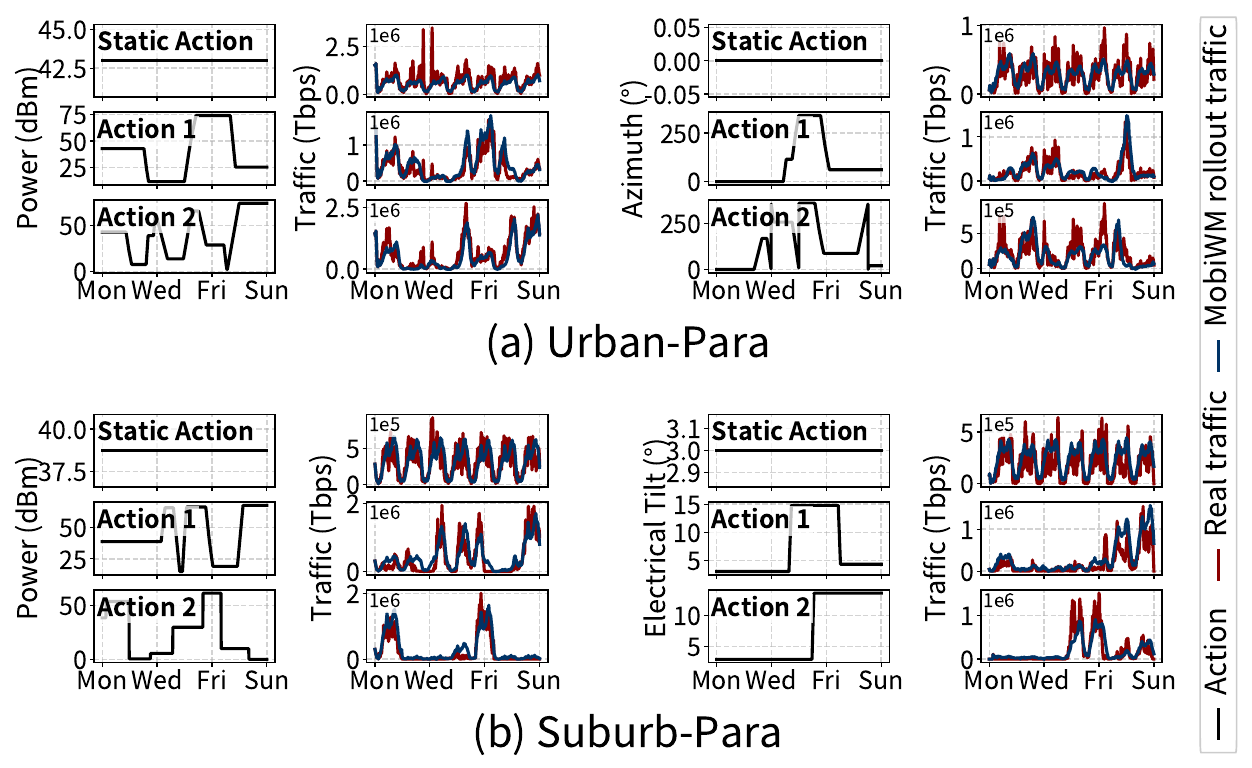}
    \caption{Evaluation of MobiWM's action-conditioned state-transition dynamics.}
    \label{fig:action}
\end{figure}

To verify the action controllability of MobiWM, we visualize the correspondence between the rollout trajectories generated by MobiWM and selected action variations in Figure~\ref{fig:action}. \emph{Static Action} represents the scenario where no action changes are applied, under which MobiWM can accurately capture the temporal fluctuations of mobile traffic. \emph{Action 1} and \emph{Action 2} correspond to applying different control actions to the same cell. The mobile traffic exhibits corresponding significant variations under these actions, and MobiWM successfully captures such dynamic responses. These results demonstrate that MobiWM effectively learns both the temporal dynamics and parameter dynamics underlying mobile traffic evolution.

\subsection{Model Efficiency}
\begin{figure}[t]
    \centering
    \includegraphics[width=0.95\linewidth]{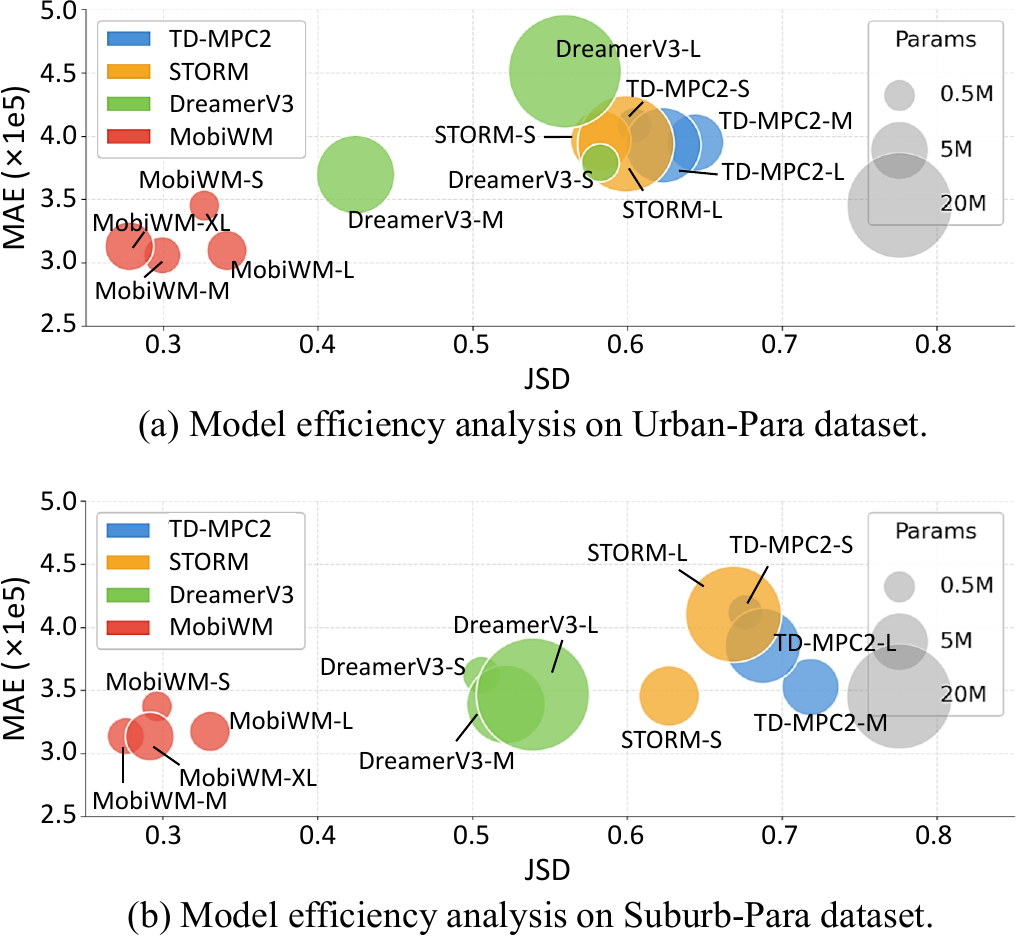}
    \caption{Model efficiency comparison. Bubble position encodes rollout accuracy (JSD, MAE); bubble size encodes parameter count.}
    \label{fig:efficiency}
\end{figure}
Figure~\ref{fig:efficiency} compares rollout accuracy (JSD, MAE) and parameter count for MobiWM (S/M/L/XL) and scaled variants of TD-MPC2, STORM, and DreamerV3.
MobiWM variants achieve lower JSD and MAE with fewer parameters in the plotted comparisons, indicating favorable parameter efficiency relative to the evaluated generic world models. Increasing model size does not monotonically improve performance: DreamerV3-L underperforms DreamerV3-M, TD-MPC2-L brings little gain, and MobiWM-XL only marginally improves over MobiWM-M. These observations suggest that architecture-specific inductive biases can be at least as important as raw capacity for the evaluated mobile traffic dynamics.

\subsection{Case Study}
\label{sec:case_study}
We use MobiWM as a learned environment for network control. We study BS energy-efficiency (EE) optimization using a model-based Actor-critic. From each frozen world model's imagined rollouts, a RewardNet estimates EE and an ActionPriorNet learns per-cell parameter adjustments that initialize and regularize a PPO actor. The optimization objective can be formulated as follows:
\begin{equation}
\label{eq:case_optim}
\begin{aligned}
    \max_{\psi}\quad \hat{\eta}_{\mathrm{EE}}
    &= \frac{\sum_{t=1}^{T}\sum_{i=1}^{N}\hat{s}_t^i}
    {\sum_{t=1}^{T}\sum_{i=1}^{N}P_{\mathrm{W}}(p_t^i)+\epsilon},\\
    \mathbf{a}_t&\sim\pi_\psi(\cdot\mid\mathbf{o}_t),\qquad
    \hat{\mathbf{s}}_{t+1}=f_\Omega(\hat{\mathbf{s}}_{\leq t},
    \mathbf{a}_{\leq t},\mathbf{c}),
\end{aligned}
\end{equation}
where $P_{\mathrm{W}}(P)=10^{(P-30)/10}$ converts dBm to Watts. We train with PPO~\cite{schulman2017ppo} and generalized advantage estimation; the reward combines rollout EE with the RewardNet estimate, and the loss includes value, entropy, and action-prior terms. Three seeds are used, with checkpoint selection and action calibration restricted to development maps.

Each fixed policy is independently replayed in Sionna on 20 terminal maps excluded from all training, selection, and calibration. From identical initial configurations and UE demand, Sionna recomputes RSRP, association/handover, traffic, and EE. We evaluate the optimization performance on \emph{Urban-Para} and \emph{Urban-Topo} datasets, and Sionna is used only for terminal evaluation. As shown in Table~\ref{tab:case}, MobiWM achieves the highest terminal EE and lowest replay MAE in both settings, demonstrating that its action-conditioned dynamics transfer reliably from imagined rollouts to the independent simulator.
\begin{table}[t]
\centering
\caption{Comparison results of EE optimization test.}
\label{tab:case}
\fontsize{8pt}{8pt}\selectfont
\setlength{\tabcolsep}{3.5pt}
\begin{tabular}{c|cc|cc}
\toprule
\multirow{3.5}{*}{\textbf{Model}} &
\multicolumn{2}{c|}{\textbf{Urban-Para}} &
\multicolumn{2}{c}{\textbf{Urban-Topo}}\\
\cmidrule(lr){2-3} \cmidrule(lr){4-5}
& \tabincell{c}{MAE\\($\times 1\text{e}5$)}
& \tabincell{c}{EE\\($\times 1\text{e}4$)}
& \tabincell{c}{MAE\\($\times 1\text{e}5$)}
& \tabincell{c}{EE\\($\times 1\text{e}4$)}\\
\midrule
No optimization & -- & 0.496 & -- & 0.441\\
\midrule
MobiFM-WM & 3.37 & 0.156 & \underline{2.03} & 0.764\\
iTransformer-WM & 6.66 & 0.171 & 7.60 & \underline{1.69}\\
TD-MPC2 & 3.44 & 0.162 & 2.57 & 1.66\\
DreamerV3 & \underline{3.06} & \underline{1.78} & 2.36 & 1.63\\
\midrule
\textbf{MobiWM (Ours)} & \textbf{2.55} &
\textbf{1.79} &
\textbf{1.77} &
\textbf{1.71}\\
\bottomrule
\end{tabular}
\end{table}

\section{Conclusion} 
\label{sec:con}
We present MobiWM, a world model for modelling traffic-parameter dynamics in the mobile network. Treating traffic as the state and antenna parameters as actions, MobiWM learns both the temporal dynamics and parameter dynamics of mobile traffic, and supports counterfactual rollouts over adjustment trajectories. Its Transformer backbone combines FSTBlocks and multimodal fusion schemes to capture action dependencies and urban context. Evaluations show that MobiWM achieves the best performance among baselines with at least 16.40\% average improvements. A downstream PPO Actor-critic study for EE optimization also supports MobiWM as a promising surrogate for counterfactual network optimization.

\newpage
\bibliography{ref_list}
\end{document}